\documentclass[useAMS,usenatbib,usegraphicx]{mn2e}

\title[Further deep imaging of HR 7329 A+B]{Further deep imaging of HR 7329 A ($\eta$ Tel A) and its brown 
dwarf companion B\thanks{Based on observations 
made with ESO Telescopes at the La Silla and Paranal Observatories under program IDs
65.L-0144, 083.C-0283, and 083.C-0276 (by us), and 
073.C-0075, 077.C-0438, 081.C-0687, 081.C-0519 (from archive)
and on observations made with the NASA/ESA Hubble Space Telescope (program IDs 7226 and 10487).
} } 
\author[R. Neuh\"auser et al.]{R. Neuh\"auser$^{1}$\thanks{E-mail:
rne@astro.uni-jena.de},
C. Ginski$^{1}$, T.O.B. Schmidt$^{1}$,
and M. Mugrauer$^{1}$ \\
$^{1}$Astrophysikalisches Institut, Universit\"at Jena, Schillerg\"asschen 2-3, 07745 Jena, Germany}

\begin{document}

\date{Accepted 2011 May 26. Received 2011 May 20; in original form 2010 November 10}

\pagerange{\pageref{firstpage}--\pageref{lastpage}} \pubyear{2011}

\maketitle

\label{firstpage}

\begin{abstract}
About $4 ^{\prime \prime}$ south of the young A0-type star HR 7329,
a faint companion candidate was found by Lowrance et al. (2000).
Its spectral type of M7-8 is consistent with a young brown dwarf companion.
Here, we report ten new astrometric imaging observations of the pair HR 7329 A and B,
obtained with the Hubble Space Telescope and the Very Large Telescope,
aimed at showing common proper motion with high significance
and possible orbital motion of B around A. 
With 11 yrs of epoch difference between the first and our last image, 
we can reject by more 
than 21 $\sigma$ that B
would be a non-moving background object unrelated to A.
%
%
We detect no change in position angle and small or no change in separation
($2.91 \pm 2.41$ mas/yr), so that
the orbit of HR 7329 B around A is inclined and/or eccentric
and/or the orbital motion is currently only in radial direction.
If HR 7329 B is responsible for the outer radius of the debris disk around
HR 7329 A being 24 AU, and if HR 7329 B currently is at its apastron
at 200 AU ($4.2^{\prime \prime}$ at 47.7 pc), we determine its pericenter distance to be 71 AU, 
its semi-major axis to be 136 AU, and its eccentricity to be $e=0.47$.
From the magnitude differences between HR 7329 A and B and the 2MASS magnitudes for the HR 7329 A+B system,
we can estimate the magnitudes of HR 7329 B
(J=$12.06 \pm 0.19$, H=$11.75 \pm 0.10$, K$_{\rm s}$=$11.6 \pm 0.1$, L=$11.1 \pm 0.2$ mag)
and then, with a few otherwise known parameters, its luminosity and mass
(20-50 Jupiter masses).
In the deepest images available, we did not detect any additional companion candidates 
up to $\le 9 ^{\prime \prime}$, but determine upper limits in the planetary mass regime.
\end{abstract}


\section{Introduction}

The A0-type star HR 7329 
(also called $\eta$ Tel or HD 181296,
distance $47.7 \pm 1.5$ pc at position
$\alpha = 19^{h} 22^{m} 51.2^{s}$ and $\delta = -54^{\circ} 25^{\prime} 26.1^{\prime \prime}$
for J2000.0 according to Perryman et al. (1997), V=5.0 mag according to Simbad)
is a member of the
$\beta$ Pic moving group 
(Zuckerman et al. 2001) with a probability of $95~\%$ (Torres et al. 2006)
to $100~\%$ (Torres et al. 2008);
it has 
%
%
an age of $\sim 12$ Myrs
(Zuckerman et al. 2001, Ortega et al. 2002, 2004, Song et al. 2003);
see Torres et al. (2008) for a review about this association.

Lowrance et al. (2000, henceforth L00) discovered a 6 mag fainter companion candidate
$\sim 4 ^{\prime \prime}$ south of HR 7329
with coronagraphic images using NICMOS at the Hubble Space Telescope (HST)
and also obtained a spectral type of M7-8; if bound to HR 7329 A,
it would be a brown dwarf companion (L00).
Guenther et al. 2001 (G01) confirmed the spectral type with an
infrared (IR) H-band spectrum obtained with the Infrared Spectrograph
and Array Camera (ISAAC) at the European Southern Observatory (ESO)
8.2m Very Large Telescope (VLT) Antu (Unit Telescope 1, UT 1),
in April 2000. Using the acquisition image,
they measured separation and position angle (PA) between HR 7329 A and B,
trying to show common proper motion, but the significance did
not exceed $\sim 1 \sigma$ (G01).

HR 7329 is one of the youngest stars known with both
a sub-stellar companion candidate 
and a debris disk (Backman \& Paresce 1993, Smith et al. 2009).
HR 7329 also has a wide stellar companion ($\sim 7^{\prime}$ separation), 
namely the F6-type star HD 181327 
with common proper motion and significant Lithium abundance (Torres et al. 2006),
which also has a spatially resolved debris disk seen in
scattered light (Schneider et al. 2006).

\section{Motivation for new deep imaging}

Torres et al. (2008) wrote in their $\beta$ Pic review
about the brown dwarf companion to be studied here (HR 7329 B = HD 181296 B):
{\em Only one of the proposed members, the brown dwarf
HD 181296 B, has no kinematical data published and its
membership can not be determined}. Such kinematic data
can be obtained by new deep high-angular resolution imaging.

L00 estimate the probability for a foreground main-sequence (MS) M7.5-type
object with H=$11.9 \pm 0.1$ mag (like the sub-stellar companion
candidate now known as HR 7329 B) to lie within $4^{\prime \prime}$
of the primary star to be $\sim 10^{-7}$ (L00). This is the probability
to find one such faint object within that separation around {\em one} target.
Since they (PI Becklin, HST program 7226) 
observed 45 stars (Lowrance et al. 2005),
the probability to find one such faint object (as MS foreground star) 
within that separation around any 
one of 45 stars is then $\sim 45 \times 10^{-7}$. 
%
%

In 2MASS, there are 4307497 sources
between K=11.5 and 11.7 mag (i.e. as faint as HR 7329 B),
so that the probability to find one such faint object in a circle
with $4^{\prime \prime}$ radius (the separation between HR 7329 A and B)
around any one of 45 stars 
is $\sim 1.8~\%$.
%
%
%

Given that HR 7239 A is a member of
the $\beta$ Pic moving group, there can be many
young stellar and sub-stellar objects in this area of the sky,
all showing a similar proper motion as members of the moving group.
Hence, the two objects HR 7329 A and the companion candidate,
even though with spectral type M7-8 and located within $\sim 4 ^{\prime \prime}$,
can in principle be two independent members of the $\beta$ Pic moving group,
which are not orbiting each other, but have distances different by a few or more pc.
In the $\beta$ Pic group the dispersion in space velocities 
is 2.0, 1.2 and 1.5 km/s in U, V, and W velocities,
respectively, from 36 member stars (Zuckerman et al. 2001, Song et al. 2003,
Moor et al. 2006, Kiss et al. 2010, Rice et al. 2010),
i.e. quite typical for a young stellar association (Jones \& Herbig 1979).

There are a total of 69 stellar members known in the $\beta$ Pic moving
group, with multiple stars counted multiple, down to spectral type
M8.5 (Zuckerman et al. 2001, Song et al. 2003, Moor et al. 2006, 
Lepine et al. 2009, da Silva et al. 2009, Schlieder et al. 2010, 
Kiss et al. 2010, Rice et al. 2010) 
- counted without the three sub-stellar companions.
These members are spread over an area 
of $\sim 2 \cdot 10^{4}$ square degrees,
i.e. almost half the sky. 
Even if we assume that there are three times more members when including all late M-type stars, brown dwarfs, 
and other as yet unknown members\footnote{Tetzlaff et al. (2011) catalogued young Hipparcos runaway 
star candidates. They investigated not only the magnitude of the velocities, but also the velocity 
vectors for each star compared to neighbouring stars and surrounding associations including the 
$\beta$ Pictoris moving group ($\beta$ Pic-Cap). In the Tetzlaff et al. catalogue, there are 112 stars 
for which the 3D velocity vector deviates from the mean motion vector of $\beta$ Pic-Cap and 
123 stars for which the 2D velocity vector does so (59 stars in both samples). Since $\beta$ Pic-Cap is 
large (radius $\sim 56$ pc), the number of stars with different kinematics that are inside the nominal 
boundaries of the group just by chance is probably large. It is thus more conservative to choose those candidates 
that are also classified as runaway star candidates owing to their large peculiar spatial velocity 
(runaway star probability larger $50~\%$). A total of 35 stars are then currently found inside 
the $\beta$ Pic-Cap boundaries and also show large peculiar space velocities and might be considered 
potential (former) members of the moving group.}, i.e. in total 207 members, then the probability 
to find one member
within $4^{\prime \prime}$ (the separation between HR 7329 A and B) of any 
of the 206 other members is $\sim 4 \cdot 10^{-6}$, 
i.e. small.
%
%

G01 wrote about the brown dwarf HR 7329 B {\em that it is consistent with a co-moving
companion of HR 7329}, but they present only $\sim 1~\sigma$ evidence, i.e. not convincing.
Chauvin et al. (2003) wrote that they {\em confirmed the known binary system HR 7329 AB},
but did not give any details nor values nor significance about the observation on HR 7329; 
their data obtained with the ESO 3.6m telescope with the ADONIS/SHARPII AO system
are not available in the ESO archive. 
Data on separation and/or position angle would be useful for future
attempts to fit the orbit of B around A, e.g. in order to measure the mass of B.

Given that a small separation together with a large magnitude difference (as in HR 7329 A+B)
does not proof companionship\footnote{There are counter-examples known like
TWA 6 and TWA 7 with faint nearby companion 
candidates (Lowrance et al. 2001, Neuh\"auser et al. 2000, respectively)
later found to be background (Macintosh et al. 2001, Neuh\"auser et al. 2002,
respectively).}, even if the probability for chance alignment would be
negligibly small, we 
obtained new deep imaging (i) to confirm companionship in Sect. 3 (as can be expected), 
(ii) to provide data points for future attempts to fit the orbit (Table 1),
(iii) to search for orbital motion of B around A in Sect. 4,
and (iv) to search for additional fainter and/or closer 
companions in Sect. 5.

\section{Observations and data reduction}

We observed HR 7329 A and B on
2009 July 1 UT 5:19h to 5:45h in service mode
using the Adaptive Optics camera NACO (for Naos-Conica
for Nasmyth Adaptive optics system with the
COude near-infrared imager and spectrograph,
Rousset et al. 2003)
located in the VLT UT4 (Yepun) Nasmyth focus
using the $2.17 \mu$m narrow-band filter
and the S13 optics with the smallest pixels available.
We obtained 22 images each of which consisting of 120
0.3454-sec 
co-added (jittered) exposures.
Darks and flats were also obtained in the same night.

We obtained the median
of the darks in the same exposure times as the flats and
the science frames, then subtracted the corresponding median dark
from the flat and science frames, then normalized the
dark-subtracted flats, and then devided the science frames
by those flats. Finally, we shifted and co-added the images.
The data reduction was done with ESO eclipse.

We also observed the binary HIP 6445 A and B
as astrometric reference with the same set-up in the same night
from UT 7:40h to 7:46h (5 images with 86 short
(0.3454 sec) exposures co-added each, also jittering).
HIP 6445 A and B have a separation of $8.360 \pm 0.0155 ^{\prime \prime}$
at 
PA=$220.9 \pm 0.1 ^{\circ}$ 
(measured from north over east and south) at the epoch 1991.25
(Fabricius et al. 2002). We reduced these data in the same way.
The separation and PA between HIP 6445 A and B
yield the pixel scale of the S13 optics for that night.
In the error budget, we include maximum possible orbital
motion between HIP 6445 A and B between the central
Hipparcos epoch and our observation:
HIP 6445 A and B have a distance of $120 \pm 15$ pc
and spectral types F4.5V and K1V,
hence a total mass of $\sim 1.8$~M$_{\odot}$,
so that maximum possible orbital motion (for a circular orbit)
in 18.25 yrs epoch difference results in
a change of 25.9 mas (milli arc sec) in separation
and $0.28^{\circ}$ in PA.

The measurement of the separation between HIP 6445 A and B
in the NACO images 
gives $629.88 \pm 0.14$ pixels. 
Compared to the Hipparcos data, and taking into account the errors 
from orbital motion in HIP 6445 A+B,
the NACO pixel scale is $13.272 \pm 0.049$ mas/pixel
(half the error budget from possible orbital motion in HIP 6445 A and B)
and the detector orientation for those images of
$0.31 \pm 0.30 ^{\circ}$, which have to be added to all
PA measurements on uncorrected images
(two thirds of the error budget from possible orbital motion in HIP 6445 A and B).
With this pixel scale, we can
convert the separation measured between HR 7329 A and B
on the detector to $4199 \pm 31$ mas
and obtain a true, corrected PA
between A and B of $166.99 \pm 0.30 ^{\circ}$.

We retrieved all archival HST and ESO data and reduced them
in a similar way as above.
In cases, where the object HR 7329 B was located within
the PSF of HR 7329 A, we first subtracted the PSF of HR 7329 A,
before we measured the position and magnitude of HR 7329 B
(using a dedicated IDL software, A. Seifahrt, priv. comm.).
Since no astrometric standards were observed in nights, for which we
took data from archives, we used the pixel scale from the 
header (fits header keyword ESO.INS.PIXSCALE gives 13.26 mas/pix without error bar);
for the archival NACO S13 data from 2004 to 2008, we used the medians
of the error bars from seven nights from 2004 to 2007 (Neuh\"auser et al. 2008),
namely $\pm 0.053$ mas/pix 
as error bar for the pixel scale from the header.
For 2004 June, which is close to epoch 2004.3 in Table 1, 
Neuh\"auser et al. (2005) gave $13.23 \pm 0.05$ mas/pix
and Eggenberger et al. (2007) gave $13.18 \pm 0.06$ mas/pix,
both for NACO S13, so that our choice here of $13.26 \pm 0.053$ mas/pix
is consistent with both of them.
For the S27 optics with the L-band data from 2008.3 (Table 1), we use the pixel
scale from the header and twice the S13 error bar, i.e. $27.19 \pm 0.11$ mas/pixel.
For the detector orientation, we have to assume correct alignment,
but use the median alignment error of $\pm 0.291 ^{\circ}$ (Neuh\"auser et al. 2008).
We include here the H-band data with short individual integration time (0.36 sec),
where HR 7329 A is not saturated, for best position measurements,
and those with long individual integration time (25 sec),
where HR 7329 A is saturated for best sensitivity and dynamic range
for detecting unknown further companion candidates;
during the observations with the 
25-sec 
individual integration times,
the field rotated from image to image, so that we de-rotated all images
before adding them up.

The HST data were reduced in the following way:
HR 7329 was observed with HST NICMOS 2 at epoch 1998.49 and 2007.75,
both in coronagraphic mode placing HR 7329 A behind an occulting hole.
After retrieving the pre-calibrated data from the HST archive,
we subtracted the PSF of A from the images to attenuate residual speckle noise.
For the 1998.49 epoch we used the two images available with a $29.9 ^{\circ}$ roll angle.
For the 2007.75 epoch we used a reference star of similar spectral type taken
with the same instrument configuration for subtraction, because there were no
rolled images available. Due to A being behind the occulting hole
of the NIC2 camera, its position could not be measured directly,
but could be calculated using its position in the acquisition image and the
telescope offset to place it behind the occulting hole given in the image headers.
We corrected the positions for geometric distortion using SMOV2 data for the 1998.49 epoch
and SMOV3b data for the 2007.75 epoch, respectively. We then calculated separation
and PA in image coordinates. Since the NIC2 pixel scale is stable since early 1997,
we derived it by averaging the measurements from late 1997 and 1998 as well as the
measurements from 2002. We took into account that the NIC2 detector is sligthly tilted towards
its focal plane, thus yielding slightly different pixel scales in x and y direction.
The results are $76.11 \pm 0.15$ mas/pix in x and $75.43 \pm 0.14$ mas/pix in y.
The orientation of the NIC2 detector y-axis was retrieved from the individual image headers.

We also observed HR 7329 with the IR imager Son OF Isaac (SofI)
at the 3.5m ESO New Technology Telescope (NTT) in 2000 (see G01) and 2009.
We reduced the data analogously as described above for NACO.
The pixel scale was determined to be $287.97 \pm 0.21$ mas/pix
and the detector orientation to be $0.02 \pm 0.04 ^{\circ}$ (to be added to direct measurements).
Before measuring the position of HR 7329 B, and hence separation and PA between HR 7329 A and B,
we subtracted the PSF of HR 7329 A with IDL.

All observations used here are listed in Table 1.

\begin{table*}
\begin{tabular}{lllllllll}
\multicolumn{9}{c} {\bf Table 1. Astrometry of HR 7329 A and B.} \\ \hline
Epoch  & Telesope,    & Band,     & exposure time              & FWHM  & separation    & PA (a)              & $\Delta$ mag      & Ref. \\
year   & instrument   & camera    & [sec]                      & [mas] & [mas]         & [$^{\circ}$]        & [mag]             & (b)  \\ \hline
1998.4931507 & HST Nicmos & F160W NIC 2 & $5 \times 143.9$            & 150 & $4170 \pm 50$ & $166.80  \pm 0.2$   & $6.9 \pm 0.1$     & L00 \\
             &            &       &                              & 152 & $4170 \pm 33$ & $166.95  \pm 0.36$  & $6.9 \pm 0.1$     & (e) \\
2000.3060109 & VLT ISAAC  & NB in K (c) & $3 \times 1.7726$      & 500 & $4097 \pm 48$ & $166.90  \pm 0.42$  & $5.7 \pm 0.1$     & G01 \\
             &            &       &                              & 476 & $4107 \pm 57$ (d) & $166.90  \pm 0.42$  & $5.8 \pm 0.1$ & (e) \\
2000.3808219 & NTT SofI   & H SF (f)  & $10 \times 46 \times 1.3$&1200 & $4310 \pm 270$& $165.8   \pm 6.7$   & (g)               & G01 \\
2004.3315068 & VLT NACO   & H S13 & $5 \times  90 \times 0.7$    & 108 & $4189 \pm 20$ & $167.32  \pm 0.22$ & $6.8 \pm 0.1$     & (h) \\
2004.3315068 & VLT NACO   & K$_{\rm s}$ S13 & $5 \times 120 \times 0.7$    &  93 & $4200 \pm 17$ & $166.85  \pm 0.22$ & $6.7 \pm 0.1$     & (h) \\
2004.3424658 & VLT NACO   & H S13 & $5 \times 180 \times 0.35$   &  78 & $4199 \pm 36$ & $167.02  \pm 0.22$ & $6.7 \pm 0.3$     & (h) \\
2004.3424658 & VLT NACO   & K$_{\rm s}$ S13 & $5 \times 120 \times 0.7$    &  76 & $4195 \pm 17$ & $166.97  \pm 0.22$ & $6.5 \pm 0.1$     & (h) \\
2006.4328767 & VLT Visir  & S IV  & $22 \times 48 \times 0.04$   & 880 & $4170 \pm 110$& $167.2   \pm 1.4$   & $6.2 \pm 0.2$     & G08 \\
2007.7534246 & HST Nicmos & F110W NIC 2 & $17 \times 160$    & 155 & $4212 \pm 33$ & $167.42  \pm 0.35$  & $7.6 \pm 0.1$     & (h)  \\
2008.3114754 & VLT NACO   & L S27 & $40 \times 150 \times 0.2$   & 165 & $4214 \pm 17$ & $166.81  \pm 0.22$ & $6.2 \pm 0.2$     & (h)  \\
2008.5983607 & VLT NACO   & H S13 & $ 3 \times  25 \times 0.36$  &  70 & $4195 \pm 17$ & $166.87  \pm 0.29$ & (g)    & (h) \\
2008.5983607 & VLT NACO   & H S13 & $27 \times 2 \times 25$      & 109 & $4194 \pm 16$& $166.20  \pm 0.29$ & (g)    & (h) \\
2009.3506849 & NTT SofI   & K$_{\rm s}$ LF (f)  & $10 \times 50 \times 1.2$&1193 & $4239 \pm 104$& $168.5   \pm 1.3  $ & (g)    & (i) \\
2009.4958904 & VLT NACO & NB in K S13 (c) & $22 \times 21  \times 0.345$ &86&$4199 \pm 31$ & $166.99  \pm 0.30$  & $6.28 \pm 0.02$   & (i) \\ \hline
\end{tabular}
Remarks: (a) Position Angle (PA) measured from north over east and south.
(b) References: L00 for Lowrance et al. (2000), G01 for Guenther et al. (2001), and G08 for Geissler et al. (2008).
(c) Narrow-Band (NB) filter inside the K-band.
(d) Separation slightly larger than in G01, because we measured the PSF center of
HR 7329 B (and hence the separation between A and B) after subtraction of the PSF of HR 7329 A from the image,
which was not done in G01.
(e) Re-reduced by us.
(f) With SofI Small Field (SF) 147 mas/pix, or SofI Large Field (LF) 288 mas/pix.
(g) HR 7329 B too faint and/or not resolved well from HR 7329 A, or HR 7329 A is saturated,
hence no useful magnitude difference obtainable.
(h) Obtained from public archive and reduced by us.
(i) Obtained and reduced by us.
\end{table*}

\section{Interpretation of astrometry}

\begin{figure}
\includegraphics[angle=0,width=1\hsize]{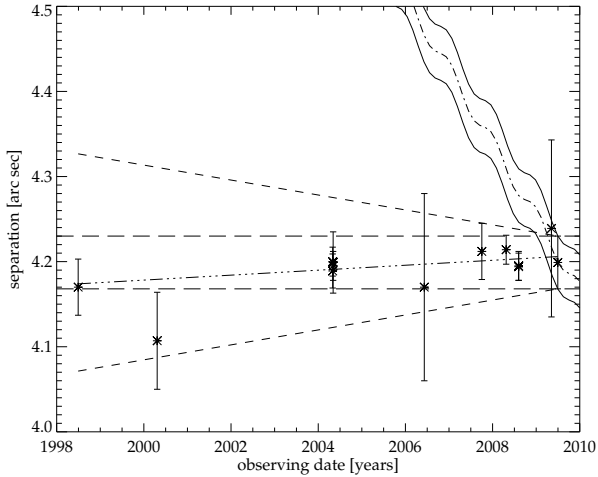}
\caption{Separation versus observing epoch
for data listed in Table 1
(except the first SofI data point due to its large error bar).
The short-dashed lines indicate maximum possible separation change due to orbital motion
for a circular edge-on orbit. The wobbled dot-dashed line is for the background
hypothesis, i.e. if HR 7329 A had moved according to its known parallactic and
proper motion, while the fainter southern object would be a non-moving object
(error cone from parallax and proper motion errors);
the data points are inconsistent with the background hypothesis
by many $\sigma$ (only the last six data points together
yield a significance of $5 \sigma$ against the background hypothesis;
the first data point alone gives a significance of 21 $\sigma$).
All data points are fully consistent with common proper motion
(within long-dashed lines).
The formal best fit for linear orbital motion yields an increase
in separation of $2.91 \pm 2.41$ mas/yr (linear dot-dashed line)
with (non-reduced) $\chi ^{2} = 3.4$ with 13 data points,
i.e. a possible (but not yet significant) marginal
detection.}
\end{figure}

\begin{figure}
\includegraphics[angle=0,width=1\hsize]{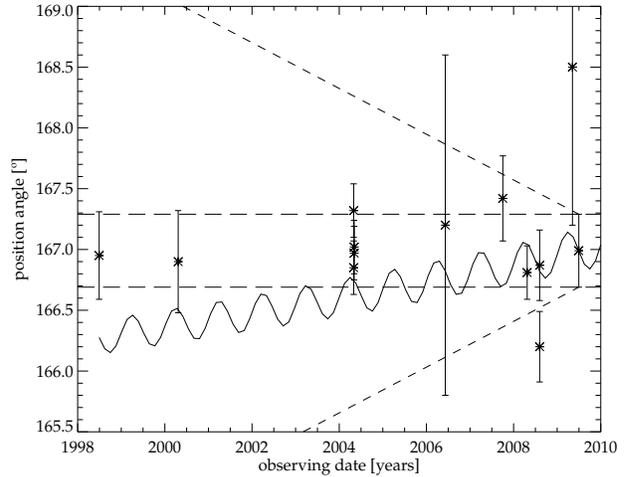}
\caption{Position angle versus observing epoch
for data listed in Table 1
(except the first SofI data point due to its large error bar).
The short-dashed lines indicate maximum PA change due to orbital motion
for a circular pole-on orbit. The wobbled line is for the background
hypothesis, which was already rejected in the previous figure
(without error cone for clarity).
All data points are fully consistent with common proper motion
(within long-dased lines).
The formal best fit for orbital motion as linear change in PA
gives only $-0.0252 \pm 0.0282 ^{\circ}$/yr
with (non-reduced) $\chi ^{2} = 12.8$ with 13 data points,
i.e. no change in PA detected.}
\end{figure}

To check for 
common proper motion and 
orbital motion, we use the astrometric data from
Hipparcos: Proper motion $\mu _{\alpha} \cdot \cos (\delta) = 25.57 \pm 0.21$ mas/yr and
$\mu _{\delta} = -82.71 \pm 0.14$ mas/yr, distance $47.7 \pm 1.5$ pc, 
both for HR 7329 A,
data from Perryman et al. (1997) and van Leeuwen (2007),
whose values differ slightly from each other, 
but are compatible within the error bars.

We show in Figs. 1 and 2 the astrometric data from Table 1,
which reject the hypothesis that HR 7329 B would
have been a non-moving background object with $\ge 21~\sigma$,
so that we can continue to regard HR 7329 A and B 
as common proper motion pair. 
We also show 
the expected maximal orbital motion
for a circular orbit of HR 7329 B around A,
being $\le 8.84$ mas/yr change 
in separation for an edge-on orbit (Fig. 1)
and $\le 0.189^{\circ}$/yr change in PA for
an pole-on orbit (Fig. 2).
We do not see any change in PA (Fig. 2) and
conclude that the (2D) orbit (on sky) is not pole-on at all.
The non-reduced $\chi ^{2}$ for the separation fit (Fig. 1)
is lower than for the PA fit (Fig. 2) - possibly indicating
that the separation errors are overestimated,
while PA errors are not overestimated. This could
be due to radial NACO detector distortions,
which are negligible for PA values.
Since the separation of the astrometric
calibrator is higher than for the HR 7329 system,
increased errors are introduced into the
radial error budget (separation errors) by jittering, 
which has to be done in order to remove the background 
in the IR in both cases.
The possibly detected change in separation due to orbital motion
($2.91 \pm 2.41$ mas/yr) is much smaller than the expected
maximum orbital motion for a circular edge-on orbit (Fig. 1),
so that we conclude that the (2D) orbit (on sky) is not a circular
edge-on orbit, but inclined and/or eccentric with HR 7329 B
currently near the apastron, hence the small motion on sky.
The orbital plane of HR 7329 B could be in the line of sight
(edge-on like the debris disk around HR 7329 A), 
with HR 7329 B currently near the largest angular separation from HR 7329 A,
but with orbital motion mostly in the radial direction;
such orbital motion could be detectable with a high-resolution
spectrum.

The possible detection of a change in the separation can also
be interpreted as evidence for slightly different proper motion
between HR 7329 A and companion candidate, namely a difference
of $0.66 \pm 0.57$ km/s (namely $2.91 \pm 2.41$ mas/yr in $47.7 \pm 1.5$ pc). 
This value is comparable to the typical
velocity dispersion in $\beta$ Pic and other young associations (see Sect. 2), 
so that one can still not exclude that the two objects called HR 7329 A and B
are two independent members of the $\beta$ Pic association.
The probability for this possibility is very low.
Orbital motion with curvature is not yet detected.

The most precise measurement of the separation between A and B is
the long 22.5 min exposure in Aug 2008 with NACO ($4194 \pm 16$ mas),
even though HR 7329 A is
saturated\footnote{Comparing the first NACO H-band S13 image from
2004 (not saturated, FWHM 108 mas) with the long exposure NACO H-band S13 image from
2008 (saturated, FWHM 109 mas) shows that the precision is similar, i.e. that
mild saturation does not matter much (see Table 1).};
the position of A was determined with MIDAS center/moment,
the position of B after subtraction of the PSF of A with MIDAS center/gauss;
with a distance of $47.7 \pm 1.5$ pc, the projected physical separation
between HR 7329 A and B is then $200 \pm 16$~AU,
the semi-major axis
for a pole-on circular orbit;
for a uniform eccentricity distribution (e=0 to 1) and a random viewing angle,
we correct this value by a factor of $1.10 ^{+0.91} _{-0.36}$ (Torres 1999, Allers et al. 2009)
and obtain $220 ^{+214} _{-84}$~AU.
We use $2.2 \pm 0.1$~M$_{\odot}$ as mass of the A0-type star HR 7329
(Tetzlaff et al. 2011)
and as total mass of HR 7329 A+B.
For this system mass, the orbital period would then
be $\sim 1900$ yrs for a semi-major axis of 200 AU
(2200 yrs for 220 AU, 345 to 6100 yrs for 64 to 434 AU).

Smith et al. (2009) directly detected the debris disk around HR 7329 A with an outer
radius being 24 AU. From the very existence (and, hence, stability) of this debris 
disk (and its outer radius), we can constrain the eccentricity of HR 7329 B even 
further: Its eccentricity cannot be too large, otherwise it would fly through the disk. 
Our deep imaging (Fig. 3 below) shows that there is no additional companion outside 
of 24 AU (or between 24 and 200 AU) with a mass larger than $\sim 20$~M$_{jup}$.
If we further assume that HR 7329 B is responsible for shaping the debris disk and
thereby fixing its outer radius, we can constrain the eccentricity as follows:
Following Pichardo et al. (2005), for the masses given here for HR 7329 A and B,
and assuming that HR 7329 B has its apocenter at 200 AU (see above) and
is responsible for the outer disk radius at 24 AU,
we determine the pericenter distance of HR 7329 B to be 71 AU, its semi-major axis
to be 136 AU, and, hence, its eccentricity to be $e=0.47$.
Then, the orbital period would be $\sim 10^{3}$ yrs.

%

\section{Deep imaging and limits on further companions}

In the three deepest images (1998 HST NICMOS, 2009 VLT NACO, and 2008 VLT NACO H-band),
no additional companion candidates were detected up to $\le 9 ^{\prime \prime}$ separation.
Companions with 12 Myrs age with 13 Jup (or 1 Jup) masses,
would have a luminosity of $\log (L/L_{\odot}) \simeq -4$ (or -5.9) (Burrows et al. 1997),
hence a magnitude difference of $\sim 10$ mag (or 14.7 mag) to HR 7329 A,
they would just be detectable at $\ge 1 ^{\prime \prime}$ (or $\ge 3 ^{\prime \prime}$,
respectively) with NICMOS and NACO (Fig. 3).
At $\sim 10 ^{\prime \prime}$ separation ($\sim 500$ AU), two companion candidates
are detected in the HST images with J=21 and H=17.5 mag,
which would be in the planetary mass regime, but probably are background;
they are outside the NACO S13 field ($\sim 9 ^{\prime \prime}$ radius around HR 7329 A),
too faint and/or blue for the NACO L-band L27 field ($\Delta L \simeq 10$ mag),
and too close and/or faint for the ISAAC and SofI images, and are, hence, detected only once.

We determined the dynamic range for all images by measuring the $3 \sigma$
level\footnote{choice confirmed by inserting and
retrieving simulated companions at this contrast level (Haase 2009)}
above the background noise for any pixel (or group of 3 or 9 or 49 pixels)
in all co-added images and compared this background flux to the flux of the
central star HR 7329 A.
The flux ratio between background and HR 7329 A is
plotted in Fig. 3 for the images with the best dynamic ranges,
i.e. where the closest and faintest companions could be detected.

\begin{figure}
\includegraphics[angle=270,width=1\hsize]{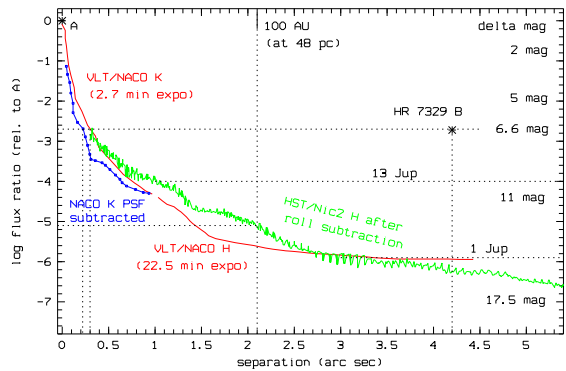}
\caption{Dynamic range plotted as log of flux ratio between background and HR 7329 A
versus separation from HR 7329 A in arc sec.
The drizzled (green) curve is for HST/NICMOS (1998) with coronagraph and roll subtraction.
The full (red) curves are for our 2009 VLT/NACO image
(inside $1 ^{\prime \prime}$ for our 2.7 min K$_{\rm s}$-band image and
outside $1 ^{\prime \prime}$ for the 22.5 min archival H-band image, where HR 7329 A is saturated),
the upper full curve without PSF subtraction and
the lower full curve (blue with dots) after PSF subtraction of HR 7329 A.
We also plot HR 7329 A and B as crosses.
Objects below the full lines cannot be detected.
Outside of $3 ^{\prime \prime}$ for VLT ($5 ^{\prime \prime}$ for HST),
the background level remains constant at
about $\Delta$H = 15 mag for VLT ($\Delta$H = 17.5 mag for HST).
At 100 AU ($2.1 ^{\prime \prime}$ at 48 pc), we could detect companions with
12.5 mag difference with HST/NICMOS
and even $\Delta$H = 14.8 mag with VLT/NACO;
other companions as faint as HR 7329 B could be detected
for $\ge 0.3 ^{\prime \prime}$ or 14 AU) with HST,
also at $\ge 0.2 ^{\prime \prime}$ with NACO after PSF subtraction.
Companions with 13 Jup masses would have 10 mag difference;
they would be detectable at $\ge 1^{\prime \prime}$ with
VLT and HST. (color figure online)}
\end{figure}

\section{Conclusions}

By several new images of HR 7329 A and B
obtained with HST/NICMOS and VLT/NACO with 11 yrs epoch difference,
we could 
reject ($\ge 21 \sigma$)
the background hypothesis, that HR 7329 B would have
been a non-moving background object unrelated to HR 7329 A.
Hence, 
HR 7329 A and B form a common proper motion pair.
The possible detection of a small linear change in separation (but no change in PA)
is consistent with an on-sky 2D orbit of B around A, which is eccentric and/or inclined.
Curvature in orbital motion
as acceleration or deceleration
would be a final proof for being gravitational bound,
but is not yet detected, as in all other
sub-stellar companions detected by direct imaging,
except PZ Tel B (Mugrauer et al. 2010).

The magnitude difference between HR 7329 A and B
is $\Delta$H = $6.75 \pm 0.10$ mag and $\Delta$K$_{\rm s}$ = $6.6 \pm 0.1$ mag (Table 1),
with K=$5.008 \pm 0.033$ mag for HR 7329 A+B (2MASS),
we get K$_{\rm s}$=$11.6 \pm 0.1$ mag for HR 7329 B;
we obtain L=$11.1 \pm 0.2$ mag for HR 7329 B (from Table 1
with L=5.0 mag for HR 7329 as A0-type star with J=H=K=L=5.0 mag);
from the magnitude difference between HR 7329 A and B in the HST F110W filter
(Table 1); we get J = $12.06 \pm 0.19$ mag for HR 7329 B,
calibrated with the M9.5 dwarf BRI B0021-02 from Persson et al. (1998)
and the NIC web site.
Those JHKL colors are consistent with spectral type M7-8 for HR 7329 B.
With a bolometric correction of B.C.$_{\rm K} = 3.10 \pm 0.05$ mag
(for M7-8, Golimowski et al. 2004), and the distance towards HR 7329 A,
we get a luminosity of $\log (L_{\rm bol}/L_{\odot}) = -2.627 \pm 0.087$ for HR 7329 B.

For 
T$_{\rm eff}$
= 2500-2800 K (for M7-8, Golimowski et al. 2004
and Luhmann 1999 intermediate scale) at $\sim 12$ Myrs,
we then derive the mass of HR 7329 B from evolutionary tracks
to be 20 to 50 Jup masses (Burrows et al. 1997, Chabrier et al. 2000, Baraffe et al. 2002).
Hence, HR 7329 B is indeed a brown dwarf.
No additional companion candidates were detected up to $\le 9 ^{\prime \prime}$.
The HR 7329 / HD 181327 system is therefore a triple system with two
stars with debris disks (HR 7329 and its wide companion HD 181327, 
Backman \& Paresce 1993, Smith et al. 2009, Schneider et al. 2006)
plus one brown dwarf (HR 7329 B).
With $\beta$ Pic (Smith \& Terrile 1984, Lagrange et al. 2010)
and PZ Tel (Smith et al. 2009, Biller et al. 2010, Mugrauer et al. 2010)
there are two more members of the $\beta$ Pic moving group,
which have both a debris disk and a sub-stellar companion,
indicating quite a large fraction and motivating further searches.

\section*{Acknowledgments}

We thank the ESO Paranal Team and ESO Users Support group.
RN, CG, and TOBS wish to acknowledge Deutsche Forschungsgemeinschaft (DFG)
for grant NE 515 / 30-1. We used Simbad and Vizier and archival data from the ESO, HST, and 2MASS.
We thank Andreas Seifahrt for his PSF subtraction routine written in IDL.
HST data were obtained from the data archive at the Space Telescope Institute,
which is operated by the association of Universities for Research in Astronomy, 
Inc. under the NASA contract NAS 5-26555.
We would also like to thank Alexander Krivov and Nina Tetzlaff for
valueable discussion about debris disks and runaway stars, respectively.

\end{document}